\documentstyle[11pt,newpasp,twoside,graphicx]{article}
\index{pulsars}
\index{supernova remnants}

\def\edcomment#1{\iffalse\marginpar{\raggedright\sl#1\/}\else\relax\fi}
\reversemarginpar

\begin{document}
\title{High-resolution radio polarimetry of Vela X}

\author{Douglas C.-J.\ Bock}
\affil{Radio Astronomy Laboratory, University of California at 
Berkeley, 601 Campbell Hall, Berkeley, CA 94720, USA; 
dbock@astro.berkeley.edu}
\author{Robert J.\ Sault, Douglas K.\ Milne}
\affil{CSIRO Australia Telescope National Facility, P.O.\ Box 76,
  Epping, NSW 1710, Australia}
\author{Anne J.\ Green}
\affil{School of Physics, University of
  Sydney, NSW 2006, Australia}

\begin{abstract}

  We present high-resolution 1.4 GHz Australia Telescope Compact Array
  polarimetric observations of Vela X, the pulsar wind nebula of the
  Vela SNR. We find that the linearly polarized emission is only partially
  correlated with total intensity. There are many depolarization
  features, some of which are coincident with foreground H$\alpha$
  filaments. Further study of these should provide measurements of the
  magnetic field in the remnant's shell.

%The data allow us to determine the direction of the
%  magnetic field on arcminute scales in filaments throughout the
%  nebula.  The correlation between areas of depolarized emission and
%  foreground H$\alpha$ filaments will also provide details of the
%  magnetic field in the SNR shell.

\end{abstract}

%---------------------------------------------------------------------
\section{Introduction}
%---------------------------------------------------------------------
%

Vela X is the pulsar wind nebula near the center of the Vela supernova
remnant (Bock, Turtle, \& Green 1998b). In total intensity, the radio
emission is dominated by synchrotron filaments which have no clear
optical counterpart (Bock et al.\ 1998a). The remnant's distance of
350 pc\footnote{Recent measurements of the pulsar's distance include
  an optical parallax of $294_{-50}^{+76}$ pc (Caraveo et al.\ 1991)
  and a radio parallax of $410\pm42$ pc (Legge 2001)} makes it one of
the nearest SNRs, and thus one of the easiest to study. Its age,
10,000 yr, is much larger than that of the Crab Nebula, and we can
hope to learn something about remnant evolution by comparing the
sources. High quality interferometric polarimetry now makes it
possible to study at sub-arcminute resolution the magnetic fields in
pulsar-powered nebulae, and in the Galaxy as a whole.

%---------------------------------------------------------------------
\section{Observations}
%---------------------------------------------------------------------
%

\begin{figure}
\centering
\includegraphics[height=4.86in,angle=90,bb=136 56 572 725]
{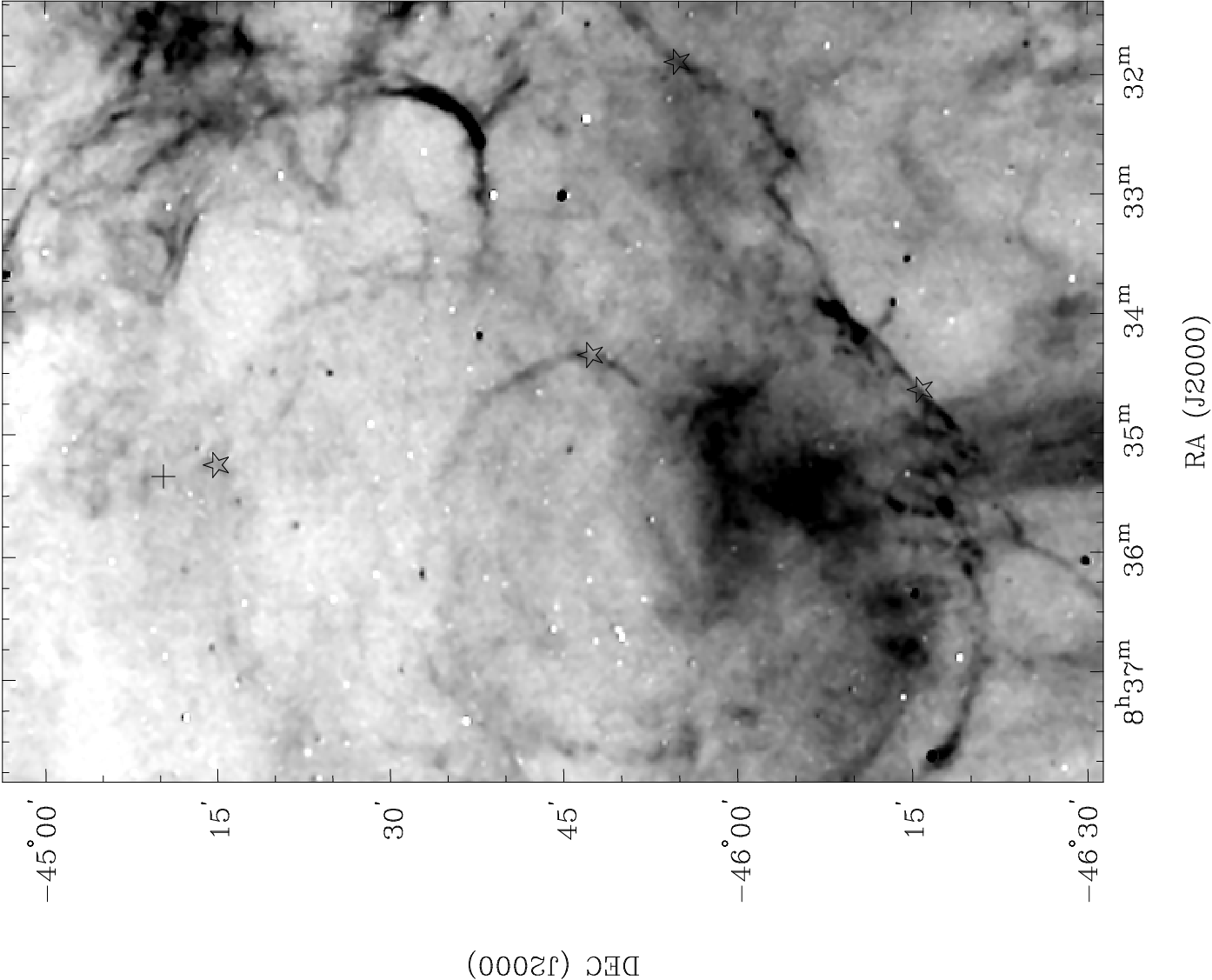}
\includegraphics[height=4.86in,angle=90]{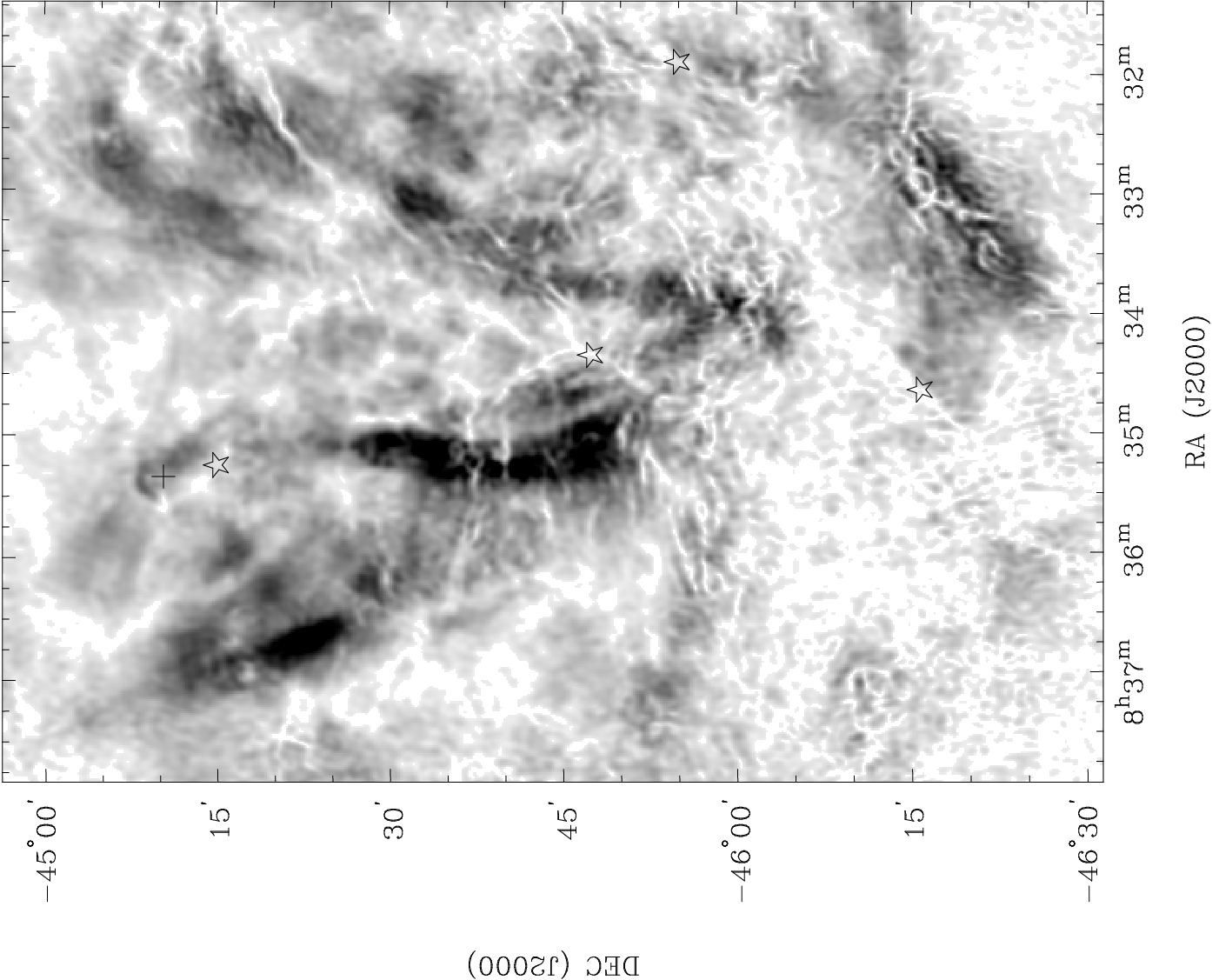}
\caption{Polarized intensity (left) and H$\alpha$ images of part of 
  Vela X. Some locations where depolarization is coincident with
  H$\alpha$ emission are marked with stars. The position of the pulsar
  is marked with a cross. The maximum polarized intensity is 180
  mJy~beam$^{-1}$. The artifacts in the H$\alpha$ image are due to
  imperfect subtraction of the stellar component.}
\label{fig:p}
\label{fig:ha}
\end{figure}

Observations of Vela X at 1.4 GHz were obtained with the Australia
Telescope Compact Array (ATCA) using configurations which yielded a
$50''\times35''$ beam. Images were made using multi-frequency
synthesis and deconvolution was performed with a maximum entropy
method algorithm. The total intensity image includes large-scale
structure which was measured with the Parkes Telescope.  However,
there is little structure in polarized intensity larger than that
sampled by the interferometer, so single-dish data has not been
included in the polarization images.  See Sault, Bock, \& Duncan
(1999) for additional details. A grayscale image showing the polarized
intensity measured from part of Vela X is shown in Figure~\ref{fig:p}.
The emission is up to 60\% linearly polarized; there is no detectable
circular polarization.  Figure~\ref{fig:ha} also shows the H$\alpha$
emission from the same region (from the survey of Buxton, Bessell, \&
Watson 1998).  Vectors representing the intensity and E-field
direction of the polarized emission are plotted on a total intensity
image in Figure~\ref{fig:ppa}.

%---------------------------------------------------------------------
\section{Discussion}
%---------------------------------------------------------------------
%
\begin{figure}
\centering
\includegraphics[height=11cm]{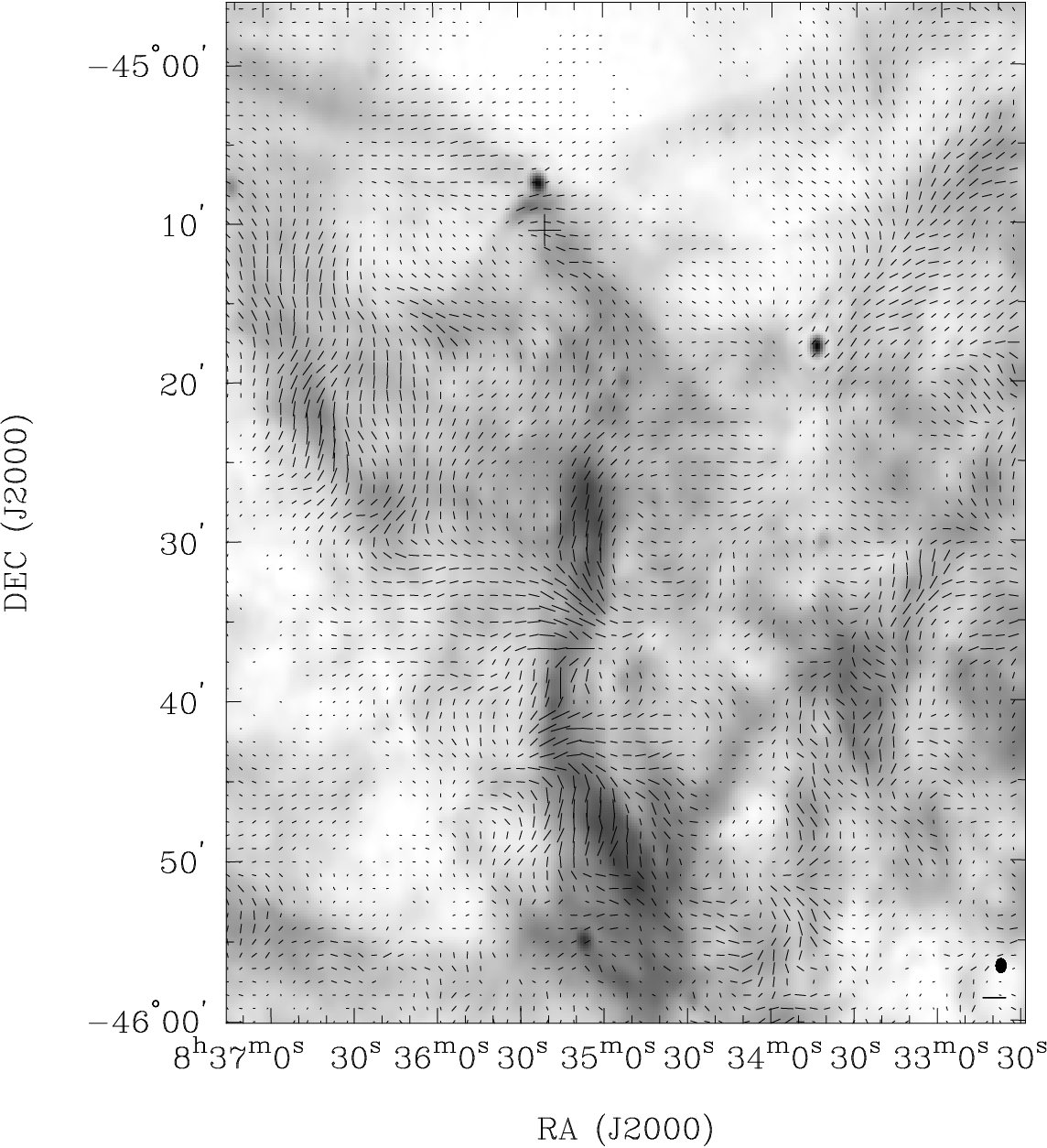}
\caption{Polarization E-vectors overlaid on total intensity. The peak
  total intensity of the extended emission is 220 mJy~beam$^{-1}$. The
  synthesized beam and a bar representing the
  maximum polarized intensity of 106 mJy~beam$^{-1}$ are shown at the
  lower right. The three unresolved sources are most likely of extragalactic
  origin.  }
\label{fig:ppa}
\end{figure}

The most striking aspect of the polarimetric image of Vela X (Figure
\ref{fig:p}) is the complex network of `canals' criss-crossing the
broader synchrotron filaments. These canals are unresolved, indicating
that they are probably due to Faraday depolarization within the
synthesized beam, caused by rapid spatial changes in the foreground rotation
measure (RM). They have no counterpart in total intensity. Similar
features have been seen in the background Galactic radiation (Gaensler
et al.\ 2001, and references therein). However, in the present case we
have been able to identify optical counterparts for many of the canals
among H$\alpha$ filaments originating in the shell of the Vela SNR
(Figure \ref{fig:ha}). By considering the geometry of the filaments
(which are probably sheets in projection) and obtaining electron
densities in the region from further optical studies, we expect to be
able to make a \emph{direct} measurement of the compressed magnetic
field in the Vela SNR shell.

The underlying more diffuse linearly polarized emission has some
overall corellation with total intensity. However, there are many
regions of disagreement. These could be either intrinsic to the source
(i.e.\ due to variations within the internal magnetic fields) or
result from depolarization by some intervening more compact region. In
the latter case, their counterpart in total intensity is the
underlying diffuse emission which forms the majority of the flux
density measured from Vela X.

The polarization E-vector direction (Figure~\ref{fig:ppa}) is a useful
diagnostic of the magnetic field. In this region, the RM is
approximately 40 rad~m$^{-2}$ (Milne 1995), indicating that the
magnetic field is generally aligned with the E-vectors plotted, and
thus lies along the filaments (the discrepant E-vectors half-way along
the central filament correspond to a region of higher RM).  We note
that the `wisp' just to the north-east of the pulsar (Bietenholz,
Frail, \& Hankins 1991) and the filament extending south do not appear
symmetric across the pulsar. Further work to make high-resolution
corrections for Faraday rotation will allow a more detailed
interpretation.

\end{document}